\documentstyle[epsf,12pt]{article}
\begin{document}


\title{\vskip-3cm
Recursion relations for two-loop self-energy diagrams on shell.}

\author{J.~Fleischer,  M.~Yu.~Kalmykov, A.~V.~Kotikov \\[2em]
{\it ~Fakult\"at f\"ur Physik, Universit\"at Bielefeld,}\\
 {\it D-33615 Bielefeld 1, Germany}
}
\date{}
\maketitle

\begin{abstract}
\noindent
A set of recurrence relations for on-shell two-loop self-energy diagrams
with one mass is presented, which allows to reduce the diagrams with 
arbitrary indices (powers of scalar propagators) to a set of the master 
integrals. The SHELL2 package is used for the calculation of special
types of diagrams. A method of calculation of higher order $\varepsilon$-
expansion of master integrals is demonstrated.
\end{abstract}

\begin{figure}[ht]
\centerline{\vbox{\epsfysize=150mm \epsfbox{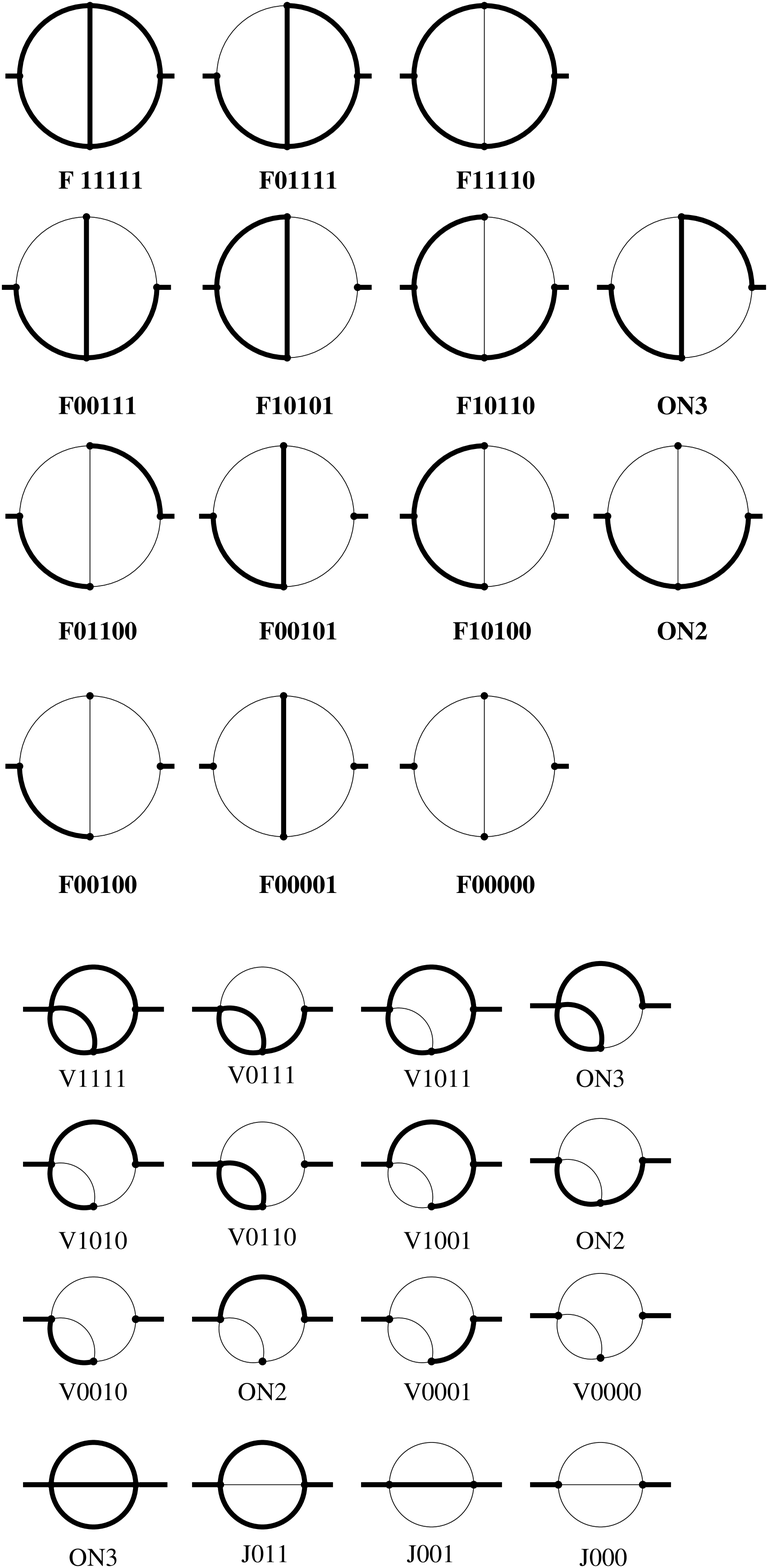}}}
\caption{\label{joint} The F, V  and J topologies.
Bold and thin lines correspond to the mass and
massless propagators, respectively.  ON3 and ON2 are diagrams
calculable by package SHELL2}
\end{figure}

Nowadays $e^+ e^-$-experiments are sensitive to multiloop radiative
corrections. The transverse part of renormalized gauge boson self-energies 
on  mass shell enter a wide class of low 
energy observables like $\Delta \rho, \Delta r$, etc. Keeping in mind 
this physical application we elaborate a FORM \cite{FORM} based 
package 
\footnote{For a review of existing packages see Ref.\cite{review}.}
\cite{our1}
for analytical calculations of on-shell two-loop self-energy diagrams
with one mass 
\footnote{This package can be used also in asymptotic expansion, 
see, e.g.\cite{asympt}.}. 
All possible diagrams of given type are shown on Fig.\ref{joint}. 
The diagrams implemented in the package SHELL2 \cite{SHELL2} 
({\bf ON3}, {\bf ON2} in our notations) and those considered in detail in 
Ref.\cite{rec} ({\bf F00000}, {\bf V0000}, {\bf J001}, {\bf J000}) 
are not discussed here. 
In contrary to  Ref.\cite{Tarasov1} we used only recurrence
relations obtained from the integration by part method \cite{rec}
without shifting the dimension of space-time. Here we apply the
triangle rule for arbitrary masses and external momentum given in 
Ref.\cite{kotikov}. We are working in
Euclidean space-time with dimension $N = 4 - 2  \varepsilon$.
The general prototype involves arbitrary integer powers of the scalar
denominators $c_L = k_L^2 + m_L^2$.
For completeness we write below our notations:
$c_1 = k_1^2 + m_1^2,~~~c_2 =  k_2^2 + m_2^2,
~~~c_3 = (k_1-p)^2+m_3^2,~~~c_4 = (k_2-p)^2+m_4^4,
~~~c_5 = (k_1-k_2)^2 + m_5^2$.
Their powers
$j_L$ are called indices of the lines. The mass-shell condition
for the external momentum now is $p^2=-m^2$.
Any scalar products of the momenta in the numerator 
are reduced to powers of the scalar propagators
(in case of V and J topologies the corresponding
lines are added). Thus, the indices may sometimes become negative.
The recurrence relations  allow to reduce all lines with negative indices
to zero and the positive indices to one or zero.

\noindent
{\bf \underline{F-topology}.}
Let us consider the diagram of F-type:

\centerline{\vbox{\epsfysize=30mm \epsfbox{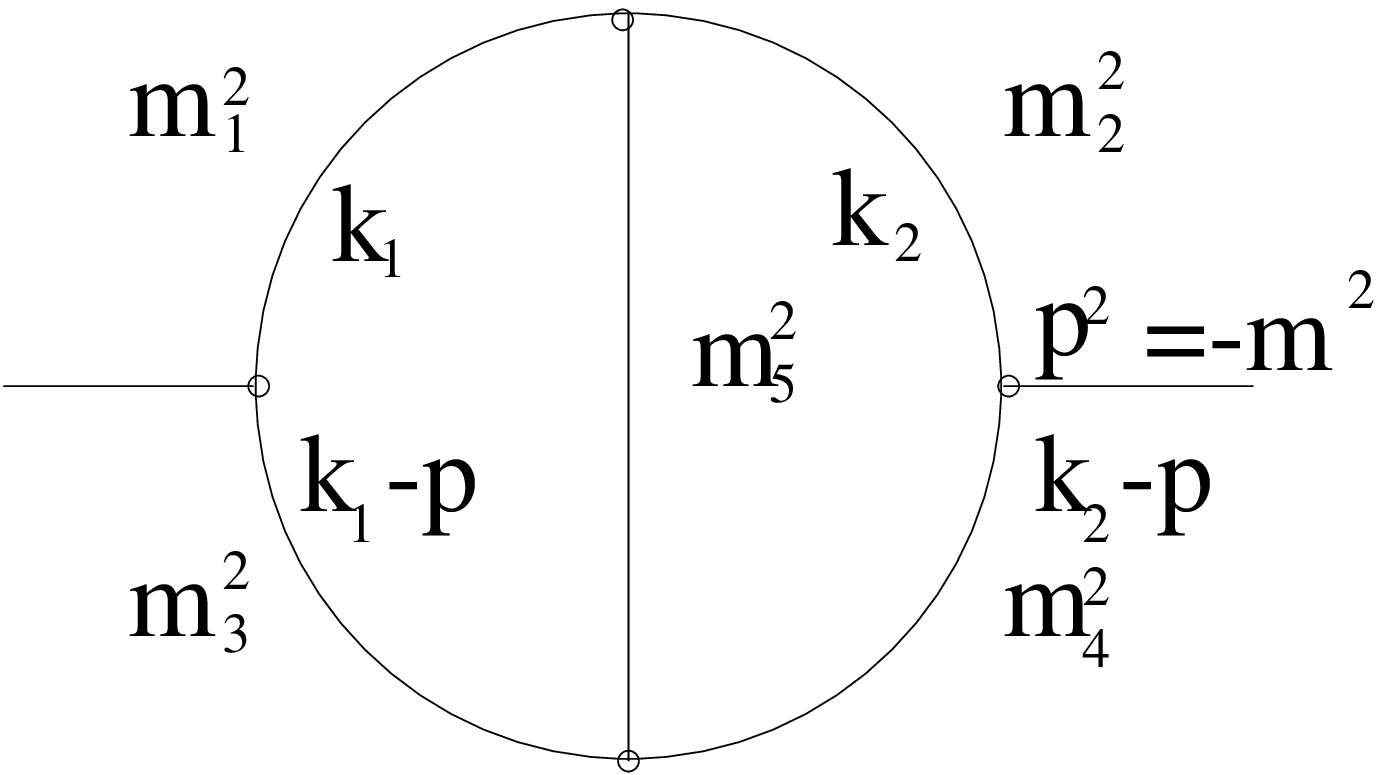}}}

\noindent
The full set of recurrence relations valid for arbitrary 
$p^2,m_1^2,m_2^2,m_3^2,m_4^2,m_5^2$ is the following:

\begin{eqnarray}
\{135\} &&  N- 2j_1-j_3-j_5 
+ \frac{j_1}{c_1} 2 m_1^2
+ \frac{j_3}{c_3} \left( m_3^2 +m_1^2 -m^2 - c_1 \right)
\nonumber \\
&& + \frac{j_5}{c_5} \left( m_5^2 +m_1^2 -m_2^2 +c_2 - c_1 \right)
 = 0, 
\nonumber \\
\{315\} &&  N - 2j_3-j_1-j_5 
+ \frac{j_3}{c_3} 2 m_3^2
+ \frac{j_1}{c_1} \left( m_3^2 +m_1^2 -m^2 - c_3 \right)
\nonumber \\
&& + \frac{j_5}{c_5} \left( m_5^2 +m_3^2 -m_4^2 +c_4 - c_3 \right)
 = 0, 
\nonumber \\
\{513\} &&  N - 2j_5-j_1-j_3 
+ \frac{j_5}{c_5} 2 m_5^2
+ \frac{j_1}{c_1} \left( m_5^2 +m_1^2 -m_2^2 +c_2 - c_5 \right)
\nonumber \\
&& + \frac{j_3}{c_3} \left( m_5^2 +m_3^2 -m_4^2 +c_4 - c_5 \right)
 = 0, 
\nonumber \\
\{245\} &&  N - 2j_2-j_4-j_5 
+ \frac{j_2}{c_2} 2 m_2^2
+ \frac{j_4}{c_4} \left( m_4^2 +m_2^2 -m^2 -c_2 \right)
\nonumber \\
&& + \frac{j_5}{c_5} \left( m_5^2 +m_2^2 -m_1^2 +c_1 - c_2 \right)
 = 0, 
\nonumber \\
\{425\} &&  N - 2j_4-j_2-j_5 
+ \frac{j_4}{c_4} 2 m_4^2
+ \frac{j_2}{c_2} \left( m_4^2 +m_2^2 -m^2 -c_4 \right)
\nonumber \\
&& + \frac{j_5}{c_5} \left( m_5^2 +m_4^2 -m_3^2 +c_3 - c_4 \right)
 = 0, 
\nonumber \\
\{524\} &&  N - 2j_5-j_2-j_4 
+ \frac{j_5}{c_5} 2 m_5^2
+ \frac{j_2}{c_2} \left( m_5^2 +m_2^2 -m_1^2 +c_1-c_5 \right)
\nonumber \\
&& + \frac{j_4}{c_4} \left( m_5^2 +m_4^2 -m_3^2 +c_3 - c_5 \right)
 = 0,
\nonumber 
\end{eqnarray}
\noindent
where both sides of these relations are understood to be multiplied by 
\newline
$
\int \frac{d^N~k_1 d^N k_2}{c_1^{j_1} c_2^{j_2} c_3^{j_3} c_4^{j_4} c_5^{j_5}}.
$
Due to the symmetry of the diagram it is sufficient to 
consider in detail only the cases $j_1 <0~~~(c_1^{|j_1|}$ in
the numerator) and $j_5 <0$. To exclude the numerator 
in the first case, we apply the following relations. 
If $j_5 \neq 1$ we solve Eq.$\{245\}$ with respect to the $\frac{j_5}{c_5} c_1$
term. If $j_2 \neq 1$ the $\frac{j_2}{c_2} c_1 $ term of $ \{524\} $ is
used. For  $j_3 \neq 1$ the linear combination  $\{245\} + \{135\}$  
is solved with respect to $\frac{j_3}{c_3} c_1.$ In case 
$j_2 = j_3 = j_5 = 1$ we apply  $\{315\}$, solved with respect to the free
term ($N-3-j_1$),
to create a denominator for which the above given relations are
applicable. The case $j_5 <0$ is considered
in detail in Refs.\cite{rec,Tarasov1,avdeev}.
In this manner F-type integrals with {\bf arbitrary}
indices are reduced to F-type integrals with only {\bf positive}
indices or V-type integrals with {\bf arbitrary} indices. 
For the former case the solution of recurrence relations is presented in 
Ref.\cite{our1}. Only eight diagrams {\bf F11111, F00111, F10101, F10110,
F01100, F00101, F10100, F00001}  with all indices equal to 1 
form the F-type basis.

\noindent 
{\bf \underline{V-topology}.}
Let us consider the V-type topology:

\centerline{\vbox{\epsfysize=30mm \epsfbox{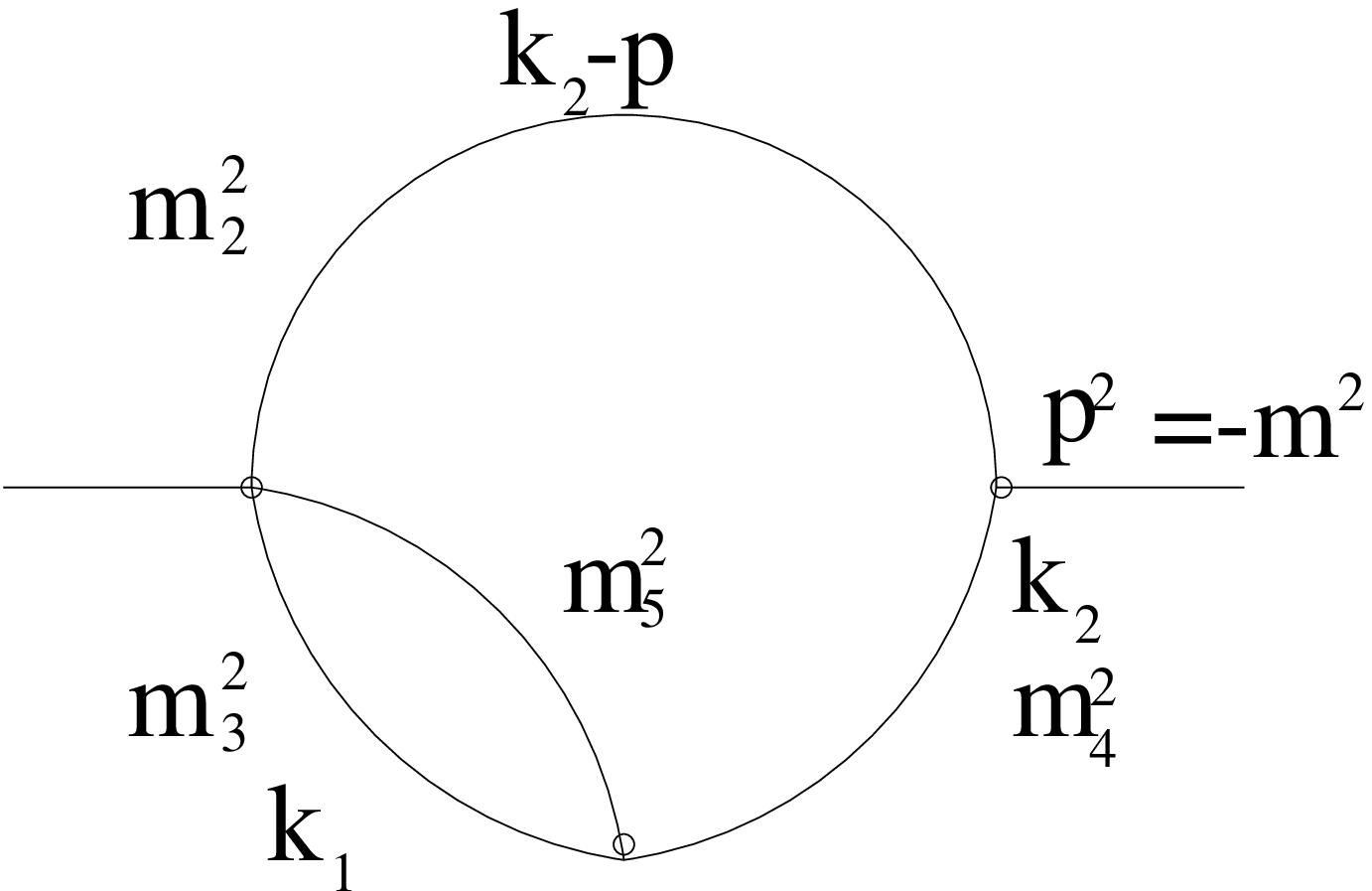}}}

\noindent
The set of recurrence relations, valid for arbitrary mass and momenta, 
consist of relation $\{ 425 \}$ and the following  ones:

\begin{eqnarray}
\{423\} && N - 2j_4-j_2-j_3
+ \frac{j_4}{c_4} 2 m_4^2
+ \frac{j_2}{c_2} \left( m_4^2 +m_2^2 -m^2 -c_4 \right)
\nonumber \\
&& + \frac{j_3}{c_3} \left( m_3^2 +m_4^2 -m_5^2 +c_5 - c_4 \right)
 = 0,
\nonumber  \\
\{530\} && N- 2j_5-j_3
+ \frac{j_5}{c_5} 2 m_5^2
+ \frac{j_3}{c_3} \left( m_3^2 +m_5^2 -m_4^2 + c_4 - c_5 \right)
 = 0,
\nonumber \\
\{350\} && N - 2j_3-j_5 
+ \frac{j_3}{c_3} 2 m_3^2
+ \frac{j_5}{c_5} \left( m_5^2 +m_3^2 -m_4^2 +c_4 - c_3 \right)
 = 0,  
\nonumber \\
\{A\} && \left( \frac{j_3}{c_3} + \frac{j_5}{c_5} \right) \left[ {\bf NUM} \right]
- \frac{j_5}{c_5}
\left( m_4^2 -m_2^2 +m^2 +c_2-c_4 \right)
 = 0,
\nonumber \\
\{B\} && 
\left( \frac{j_2}{c_2} + \frac{j_4}{c_4} + \frac{j_5}{c_5}  
\right)
\left( m_4^2 -m_2^2 +m^2 +c_2-c_4 \right)
\nonumber \\
&& 
- \frac{j_5}{c_5}  \left[ {\bf NUM} \right]
- \frac{j_2}{c_2} 2 m^2   = 0,  
\nonumber \\
\{C\} && \frac{j_2}{c_2}  \left[ {\bf NUM} \right]
+ \left( \frac{j_2}{c_2} + \frac{j_4}{c_4} + \frac{j_5}{c_5}  
\right)
\left( m_5^2 -m_3^2 - m_4^2 + c_3 + c_4 - c_5 \right)
\nonumber \\
&& 
- 2 \frac{j_5}{c_5} \left( c_3 -m^2 \right) = 0,  
\nonumber 
\end{eqnarray}
\noindent
where we introduce 
$$
\left[ {\bf NUM} \right] \equiv
\frac{
\left( m_4^2 -m_2^2 +m^2 +c_2-c_4 \right)
\left( m_5^2 -m_3^2 -m_4^2 +c_3+c_4-c_5 \right)}{2 c_6},
$$
$c_6 = c_4 - m_4^2$
and the zero-index in the above relations denotes lines with zero mass
and zero index. Let us discuss in detail the relations $\{A,B,C \}$. Due
to the presence of a four-line vertex in this case, some scalar product arising in 
the recurrence relations cannot be expressed as linear combination of denominators. 
Nevertheless this scalar product can be presented as a nonlinear 
combination of a denominator and a `new' massless propagator
(see Ref.\cite{similar}). Let us consider the following 
distribution of momenta: $c_2 =  (k_2+p)^2 + m_2^2;
c_3 = k_1^2+m_3^2;c_4 = k_2^2+m_4^4; c_5 = (k_1-k_2)^2 + m_5^2.$ 
Then relation $\{ B\}$ reads
$$
0 \equiv  \int d^N k_2 \frac{\partial}{\partial k_2^\mu} 
\left\{ \frac{p^\mu}{c_2^{j_2} c_4^{j_4} c_5^{j_5} } \right \} 
\to 2 m^2 \frac{j_2}{c_2} - 2 \frac{j_5}{c_5} k_1p
- \left( \frac{j_2}{c_2} + \frac{j_4}{c_4} +\frac{j_5}{c_5}  \right) 
k_2 p=0.
$$
\noindent 
The scalar product $k_2p$ is rewritten in the following way:
$k_2p = c_2 - c_4 + m_4 - m_2 + m^2,$
whereas $k_1 p $ can be presented by means of the projection 
operator: $k_1p = A(k_1,p,k_2) + \frac{(k_1 k_2)(p k_2)}{k_2^2}$, 
where $ A(q, r, p, ) = q^\mu
\left(\delta_{\mu \nu} - \frac{_\mu p_\nu}{p^2} \right) r^\nu $
satisfies the property that odd power of 
$A(q, r, p)$ drop out after integration. Due to this property we have 
$k_1 p = \frac{
\left( m_4^2 -m_2^2 +m^2 +c_2-c_4 \right)
\left( m_5^2 -m_3^2 -m_4^2 +c_3+c_4-c_5 \right)}{4 c_6},
$
\noindent 
where $c_6 = c_4 - m_4^2$. If $m_4^2 \neq 0$,
the expression $1\over c_4 c_6$ can be simplified by partial fraction
decomposition. 

\noindent 
{\bf \underline{Numerator for V-topology}.}
For $j_2,j_3$ or $j_5 <0$ the initial diagram can be
reduced to a two-loop tadpole-like integral by Eq.(2.10) in \cite{DT}. 
For $j_4 <0$ we apply the following relations.
If $j_5 \neq 1$ we solve Eq.$\{350 \}$ with respect to the $\frac{j_5}{c_5}
c_4$ term. If $j_3 \neq 1$ the $\frac{j_3}{c_3} c_4 $ term of $ \{530\} $ is
used. For  $j_2 \neq 1$ the linear combination  $\{425\} + \{350\}$
is solved with respect to $\frac{j_2}{c_2} c_4.$  The case $j_2 = j_3 =
j_5 = 1$ requires additional consideration. We distinguish the following
cases for {\bf on-shell} integrals:

1. $m_3^2 = m_5^2 = m^2$, ~~~~~$m_2^2 =0. $
$$
N -3j_5 = 
\frac{j_2}{c_2}(c_5-c_3 )
\left( \frac{m^2}{c_4} + 1 \right)
+ \frac{j_5}{c_5} (c_3-c_4 )
+ \left( j_2 + 2j_4 \right) \frac{c_5-c_3}{c_4} 
- \frac{j_5}{c_5} 4 m^2.  $$

2. 
$m_2^2 = m_3^2 = m^2,$~~~~~$m_5^2=0.$
$$
N -3 j_2  = 
\frac{j_5}{c_5} \frac{c_2}{c_4} m^2
- \frac{j_2}{c_2} \left( 4 m^2 + c_4 \right)
+ \left(j_5 + 2j_4 \right) \frac{c_2}{c_4}
+ \frac{j_5}{c_5} \frac{c_2}{c_4} \left( c_4 - c_3 \right). 
$$
\noindent
The other cases have been consider in Ref.\cite{SHELL2}. 
In this manner the V-type integrals with {\bf arbitrary}
indices are reduces to V-type integral with only 
{\bf positive} indices or J-type integral with {\bf arbitrary}
indices.  For the former case the solution of recurrence relations is 
presented in Ref.\cite{our1}. 
The complete set of basic integrals is just given by 
{\large \bf V1111, V1001} with all indices equal to 1.

\noindent
{\bf \underline{J-topology}.}
The integrals of this type are discussed in detail in Ref.\cite{Tarasov1}.
We only mention here that to reduce the numerator, 
Eq.(7) of \cite{Tarasov2} is used. The master integrals are
the following: one prototype {\bf J111} with all indices equal to 1, and
two integrals  of {\bf J011}-type: with indices 111 and 112, respectively.

\noindent
{\underline{\bf Master-integrals}.}
To obtain the finite part of two-loop physical results one needs
to know the finite part of the
F-type integrals, V- and J-type
integrals up to the $\varepsilon$-part, and one-loop integrals up to the
$\varepsilon^2$-part. 
The detailed discussion of the calculation of the
master-integrals up to the needed order and a comparison with earlier existing 
results is given in Ref.\cite{our2}.  
Here we present the result of
numerical investigations of the integral ${\bf J011}(1,2,2,m).$ 
The main idea is very simple \cite{Broad}: knowledge of a high precision
numerical value of the integral and a set of basic irrational numbers 
allows to find the analytical result by applying the FORTRAN program 
{\bf PSLQ} \cite{pslq}. Inspired by this idea we found
the next several orders of the $\varepsilon$-expansion of the above integral.
High precision numerical results for diagrams with smallest threshold far
from their mass shell (e.g. {\bf F11111},{\bf V1111}, {\bf J111}, {\bf J011}) 
can be obtained by the method elaborated in \cite{Pade}: Pad\'e approximants
are calculated from the small momentum Taylor expansion of the
diagram. 
The main problem in this procedure is to find the proper
basis. At the present moment we don't know the general solution of
this. Nevertheless, for {\bf J011}(1,2,2,m) diagram we guessed the
basis up to ${\cal O} (\varepsilon^5)$ with the following the result:

\begin{eqnarray}
&&
m^2 {\bf J011}(1,2,2,m) =  \frac{2}{3} \zeta_2
- \varepsilon \frac{2}{3} \zeta_3 + \varepsilon^2 3 \zeta_4
- \varepsilon^3 \left \{ 2 \zeta_5 + \frac{4}{3} \zeta_2 \zeta_3
\right\}
\nonumber \\
&&
+ \varepsilon^4 \left \{ \frac{61}{6} \zeta_6 + \frac{2}{3} \zeta_3^2
\right\}
- \varepsilon^5 \left \{ 6 \zeta_7 + 4 \zeta_2 \zeta_5 + 6 \zeta_3
\zeta_4 \right\}
+ {\cal O} (\varepsilon^6), 
\label{new}
\end{eqnarray}

\noindent
where the general factor is 
$\frac{\Gamma^2(1+\varepsilon)}{ \left(4 \pi \right)^\frac{N}{2}
\left( m^2 \right)^{2 \varepsilon}} $ is assumed. 
The ${\cal O} (\varepsilon^6)$ term is not expressible in terms of 
$\zeta$-function $(\zeta_8, \zeta_3 \zeta_5)$ only, so that 
a new irrational, like $\zeta(5,3)$ \cite{david} e.g., may arise.

\noindent
{\bf \underline{Acknowledgments}}
We are grateful to A.~Davydychev for useful discussions 
and to O.~Veretin for his help in numerical calculation.
M.K. and A.K.'s research has been supported by the DFG
project FL241/4-1  and in part by RFBR $\#$98-02-16923.

\end{document}